\documentclass[lettersize,journal]{IEEEtran}
\usepackage{amsmath,amsfonts}

\usepackage{array}
\usepackage[caption=false,font=footnotesize,labelfont=rm,textfont=rm]{subfig}

\usepackage[font=footnotesize,labelfont=rm,textfont=rm]{subfig}
\usepackage{caption}
\usepackage{textcomp}
\usepackage{stfloats}
\usepackage{url}
\usepackage{verbatim}
\usepackage{graphicx}
\usepackage{color}
\usepackage{algorithm}
\usepackage{algpseudocode}
\usepackage{amsmath}
\usepackage{float}
 \usepackage{epstopdf}
\usepackage{booktabs}
\usepackage{multirow}
\usepackage{bbding}

\usepackage[numbers,sort&compress]{natbib}

\def\BibTeX{{\rm B\kern-.05em{\sc i\kern-.025em b}\kern-.08em
    T\kern-.1667em\lower.7ex\hbox{E}\kern-.125emX}}
\usepackage{balance}

\begin{document}
\title{User-Centric Data Management in Decentralized Internet of Behaviors System}
\author{\IEEEauthorblockN{
		 Shiqi Zhang, 
            Dapeng Wu,~\IEEEmembership{Senior Member, IEEE,}
 Honggang Wang,~\IEEEmembership{Fellow, IEEE,}
and Ruyan Wang
}\\

\thanks{
This work was supported in part by the National Natural Science Foundation of China (62271096, U20A20157), Natural Science Foundation of Chongqing, China (CSTB2023NSCQ-LZX0134, CSTB2024NSCQ-LZX0124), University Innovation Research Group of Chongqing (CXQT20017), Youth Innovation Group Support Program of ICE Discipline of CQUPT (SCIE-QN-2022-04) ({\it Corresponding authors: DapengWu})

Shiqi Zhang, Dapeng Wu, and Ruyan Wang are all with School of Communications and Information Engineering, Chongqing University of Posts and Telecommunications, Chongqing 400065, China, and also with Advanced Network and Intelligent Connection Technology Key Laboratory of Chongqing Education Commission of China, Chongqing Key Laboratory of Ubiquitous Sensing and Networking. (email: d220101025@stu.cqupt.edu.cn, wudp@cqupt.edu.cn, wangry@cqupt.edu.cn)

Honggang Wang is the founding Chair and Professor of the Department of Graduate Computer Science and Engineering, Katz School of Science and Health, Yeshiva University in New York City.  (email: Honggang.wang@yu.edu )

 }}

\markboth{Journal of \LaTeX\ Class Files,~Vol.~X, No.~X, X~202X}%
{How to Use the IEEEtran \LaTeX \ Templates}

\maketitle

\begin{abstract}

The Internet of Behaviors (IoB) is an emerging concept that utilizes devices to collect human behavior and provide intelligent services. Although some research has focused on human behavior analysis and data collection within IoB, the associated security and privacy challenges remain insufficiently explored. This paper analyzes the security and privacy risks at different stages of behavioral data generating, uploading, and using, while also considering the dynamic characteristics of user activity areas. Then, we propose a blockchain-based distributed IoB data storage and sharing framework, which is categorized into sensing, processing, and management layers based on these stages. To accommodate both identity authentication and behavioral privacy, zero-knowledge proofs are used in the sensing layer to separate the correlation between behavior and identity, which is further extended to a distributed architecture for cross-domain authentication. In the processing layer, an improved consensus protocol is proposed to enhance the decision-making efficiency of distributed IoB by analyzing the geographical and computational capability of the servers. In the management layer, user permission differences and the privacy of access targets are considered. Different types of behavior are modeled as corresponding relationships between keys, and fine-grained secure access is achieved through function secret sharing. Simulation results demonstrate the effectiveness of the proposed framework in multi-scenario IoB, with average consensus and authentication times reduced by 74\% and 56\%, respectively.

\end{abstract}

\section{Introduction}

The Internet of Things (IoT) has become integrated into different aspects of human life with the rapid development of communication networks, data analytics, and hardware devices. Transforma Insights predicts that up to 2030, these devices will reach 24.1 billion and generate massive amounts of data containing personal information. Due to advances in emerging technologies such as artificial intelligence (AI) and big data analytics, this data can be explored and utilized effectively. IoT aims to achieve seamless interconnection of smart devices, enabling analysis and prediction through data exchange and sharing to optimize device operation and provide information services\cite{WW1}. Further, Göte Nyman proposed Internet of Behaviors (IoB) \cite{WW16} based on IoT, which aims to extract human behavior (hereafter referred to as behavior) from rich data. It facilitates the shift of IoT from providing information-based services to intelligent services, making smart devices no longer limited to data collection, but also in tracking and analyzing behavior. Most studies \cite{WW3,WW4,WW5,WW6,WW7} aim to leverage behavior to analyze and infer human intentions, providing potential services. Haya et al. \cite{WW4} integrated IoB with the concept of Explainable Artificial Intelligence (XAI), proposing a user behavior system. The authors in \cite{WW5} introduced a decentralized IoB framework that promotes energy sustainability by tracking, analyzing, and influencing the behaviors of IoT devices. Other research in \cite{WW6,WW7} focused on recognizing human activities through the perception of behavior using single-type and multi-type devices.

Distinct from previous studies, we focus on the security and privacy challenges in the process of behavioral data collection and use. Therefore, the behavioral processing flow in IoB needs to be explored, which consists of four main parts: (1) Behavior sensing: IoB collects behavioral data from infrastructure and mobile devices, both physical and online behavior. (2) Behavior analysis: Edge servers use AI, pattern recognition, and other technologies to extract valuable information from behavioral data. (3) Behavior inference: IoB predicts user intentions by analyzing historical and current behavioral data, enabling the behavioral network to predict the user's next behavior. (4) Behavioral services: Edge servers invoke devices to provide smart services based on behavioral inferences.

\begin{figure}[!t]
\centering
\includegraphics[width=3.0in]{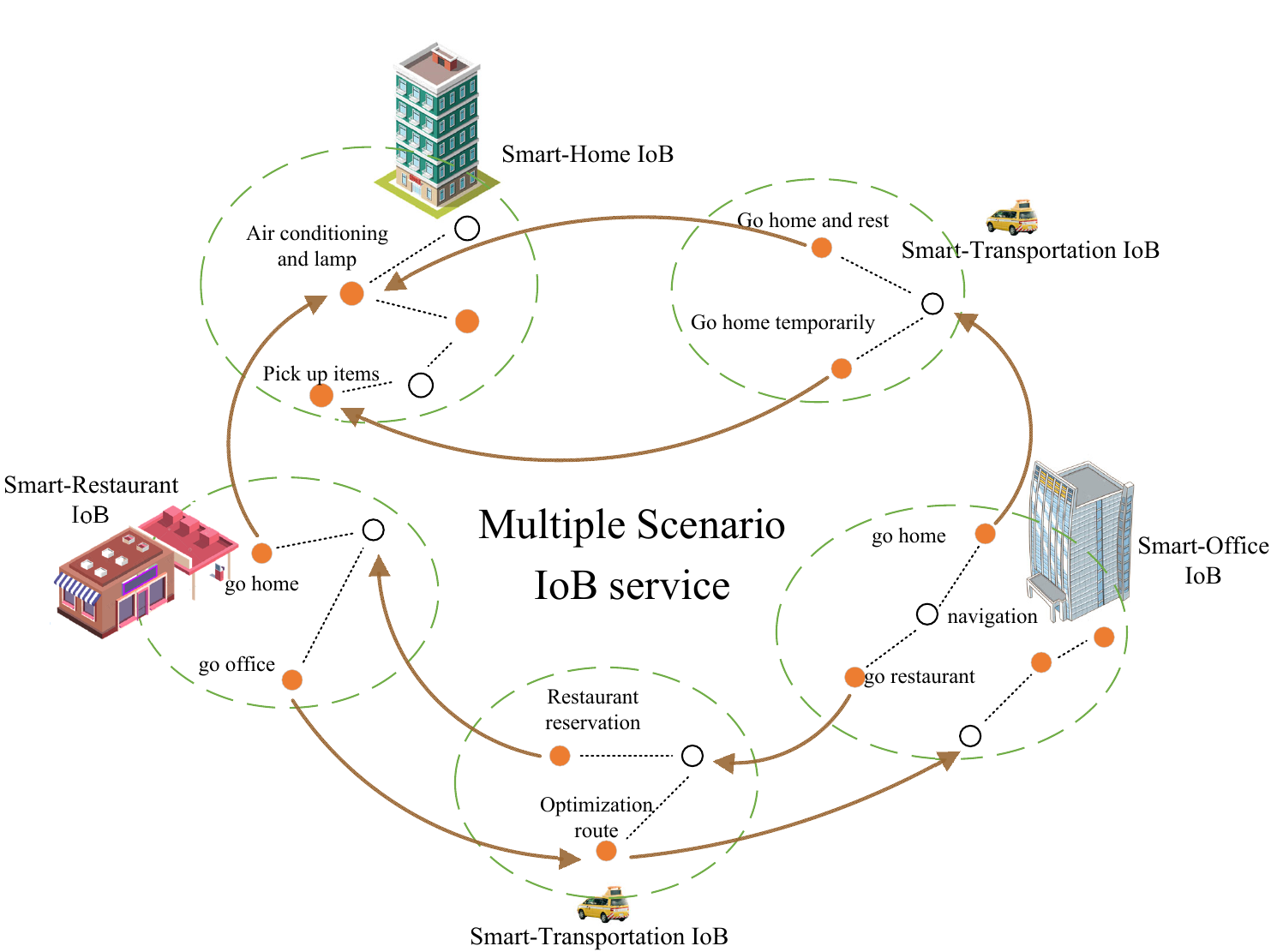}
\caption{\centering Multi-scenario IoB} 
\label{Fig1}
\end{figure}

Depending on the complexity of the scenario, IoB services are categorized into single and multi-scenario cases. An example of a single-scenario IoB is a smart home scenario that remotely wakes up the air conditioner or lights by acquiring the user's intent to come home. In multi-scenario IoB \cite{WW8}, the user activity area is a set of multiple single-scenarios such as company-restaurant-home, and the behavioral data collected in these scenarios are jointly analyzed by servers. In Fig.1, a user may set a navigation route in the office parking lot, and this behavior is uploaded to a nearby edge server. This edge server then shares the information with other servers in the network (such as RSU or the home gateway server). These edge servers infer the user’s intention based on stored behavioral data. If the server infers that the user's next behavior is to go to a restaurant, the RSU will remind the user to adjust the driving speed and make a reservation for the corresponding restaurant. The gateway server then estimates the arrival time at home based on the user's meal time or road traffic conditions and adjusts the room temperature and lighting. This example illustrates how multi-scenario IoB can effectively adapt to situations where users frequently change activity areas. By leveraging joint analysis across different areas, the system can deliver higher-quality applications (smart business and transportation). For example, the joint analysis of multiple scenarios helps the system to understand that the user's need at this moment is to arrive home early to rest but need to go to dinner. Therefore, the system will recommend fast food restaurants that do not require queuing for the user. In route selection, the system will recommend a route that is not congested but may have poor road conditions. Due to these advantages, we focus on addressing security challenges in multi-scenario IoB.

In IoB, behavioral data inherently includes sensitive user information. If this data is not protected, it will raise security and privacy concerns for users. Data security primarily involves secure data storage and sharing, preventing unauthorized users from maliciously tampering with, deleting, or accessing data. Due to the distributed nature of servers and uncertainties in the data transmission process, multi-scenario IoB faces the following additional challenges: (1) Ensuring the integrity and authenticity of behavioral data across the different scenarios. (2) Achieving consensus among edge servers for data sharing and analysis. (3) Safeguarding user privacy throughout the processes of data uploading, sharing, and accessing.

Blockchain ensures the integrity, traceability, and reliable sharing of behavior \cite{WW9}. It is one of the key solutions for addressing the challenges mentioned above. Prior studies have utilized blockchain for data storage and sharing. The authors in \cite{WW10} proposed a blockchain-based secure data sharing framework within a Multi-access Edge Computing (MEC) system to reduce energy consumption. Lu et al. \cite{WW11} designed a blockchain-enabled secure data sharing architecture for distributed systems, integrating it with federated learning to address data sharing challenges as machine learning problems. To tackle issues related to secure data storage, access control, data updates, and deletions, the authors in \cite{WW12} proposed a blockchain-enhanced access control scheme supporting traceability and revocability. Different from the above perspective, we focus on achieving rapid consensus and storage of single-user data across multiple servers, ensuring that edge servers in various areas can perform real-time data storage and analysis.

Due to the dynamic nature of the user activity area, it is crucial to address the validity of data consensus and security issues related in multi-scenario. To tackle these challenges, we propose a distributed IoB data storage and sharing framework that integrates IoB and blockchain. This framework leverages edge collaboration to ensure the accuracy and integrity of data. For multi-scenario data analysis, receiving edge servers create behavior blocks and utilize a consensus protocol to facilitate communication among all edge servers. The consensus protocol is designed by clustering edge servers based on geographical and computational differences, enhancing the framework's efficiency and adaptability.

Data privacy in IoB consists of two aspects, on the one hand, the behavior of establishing correspondence with the user's identity leads to the disclosure of the user's privacy. To resolve the conflict between identity legitimacy and behavioral authenticity, it is essential to diminish the strong association between identity and behavior. 

We consider that users create credentials based on their identities, enabling blockchain nodes to verify the legitimacy of these identities without disclosing personal information. This approach uses zero-knowledge proof(ZKP) \cite{WW13}, where the prover (user) proves their identity to the verifier (blockchain node) by credential, which credential is guaranteed to be secure by computational hardness assumption, without revealing personal information. Further, we enhance the ZKP protocol through non-interactive operations and cross-domain mechanisms aiming to increase the dynamics and efficiency of the framework, and integrate the solution into distributed IoB.

On the other hand, data privacy includes access control and verification of access rights, which requires that users have access to data with appropriate rights and that the access targets are not leaked. Therefore, this paper designs a privacy-preserving access control mechanism in a distributed system where different types of behaviors are modeled as correspondences between keys through Function Secret Sharing(FSS), and servers in the blockchain jointly verify the permissions through the keys and are unaware of the specific target of the access. The contributions of this paper are as follows:

\begin{list}{}{}
\item $\bullet$ {We propose a blockchain-based distributed IoB data storage and sharing framework to enable secure storage and sharing of behavioral data across multi-scenario. Specifically, the framework is divided into three layers: the behavior sensing layer, the behavior processing layer, and the behavior management layer.}
\item $\bullet$ {We study the separability of identity and behavior in the sensing layer and propose a non-interactive ZKP cross-domain authentication scheme. This scheme addresses the privacy issues caused by the strong association between behavior and identity, ensuring user legitimacy and privacy while enabling efficient and cross-domain authentication.}
\item $\bullet$ {In the processing layer, we consider the geographic and computational differences of edge servers to implement node cluster management through clustering. The consensus algorithms within and across clusters are optimized to improve the dynamism and effectiveness of the proposed framework.}
\item $\bullet$ {In the management layer, we establish a privacy-preserving personalized access control mechanism based on FSS. This mechanism models access rights as a one-to-one mapping between functions and keys. Multiple nodes verify a user's access rights to behavioral data through FSS without revealing the target of access.}
\end{list} 

\section{Architecture of User-Centric Decentralized Internet of Behaviors}
Fig.2 shows a multi-scenario IoB in which the user's work or lifestyle is dynamic and will move in and out of different areas or leave an area for a long time. For example, when a user leaves a company or develops a new sports interest, servers in the relevant area join or leave the IoB. We propose a blockchain-based distributed IoB, where edge servers from multiple scenarios serve as nodes (hereinafter referred to as node) within the blockchain. The framework optimizes the node topology, communication, and management methods among the nodes to ensure the effectiveness of the consensus protocol after the increase of nodes and improve the scalability. The framework is divided into three layers:

\begin{figure*}[!t]
\centering
\includegraphics[width=4.0in]{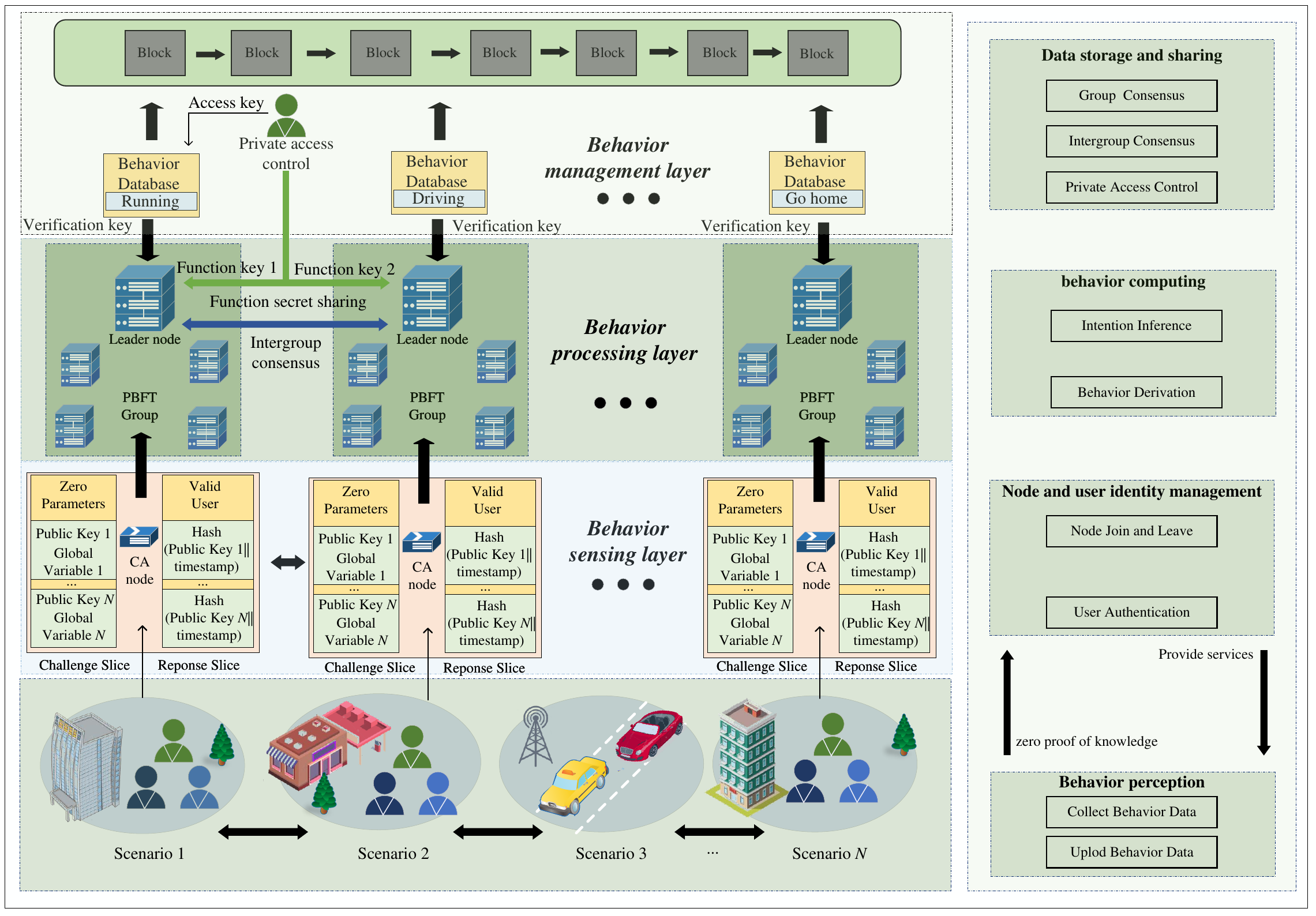}
\caption{System model}
\label{Fig2}
\end{figure*}

$\text{\textbf{Behavior Sensing Layer:}}$ This layer verifies user legitimacy, and collects behavioral data. Before collecting data, nodes authenticate user identities to ensure the association between behavior and identity, along with all potential relationships, thereby guaranteeing the accuracy of services. We propose an across-domain authentication scheme to verify user legitimacy, enabling users to transition quickly between different scenarios.

$\text{\textbf{Behavior Processing Layer:}}$ This layer ensures that each node agrees on the behavioral data through consensus. Specifically, behavioral data is required to be distributed to nodes in different scenarios to jointly perform behavioral analysis or establish behavioral networks in the IoB. So, the node that receives the data first initiates a consensus protocol to ensure that all nodes receive the same behavior and then generates a new block to store that data. We cluster the nodes by considering different node computational capabilities and geographic locations and design different consensus protocols within and between groups to improve efficiency.

$\text{\textbf{Behavior Management Layer:}}$ This layer is responsible for behavior storage and access. We achieve behavioral data access and sharing for users with different privileges based on FSS. FSS maps different behavioral databases to various function outputs, establishing a one-to-one link between authentication keys and access keys. Permissions are jointly verified by nodes, allowing users to access specific behavior according to their permissions while protecting the access target.

\section{Behavior sensing layer}
\noindent

To ensure that behavior corresponds to users, nodes storing behavioral data must verify user identities. This verification establishes a connection between behaviors and user identities, potentially compromising privacy. Thus, the conflict between identity legitimacy and privacy must be resolved before the behavior of collection. This paper addresses this issue by decoupling behavior from identity and integrating a ZKP protocol, enabling users to authenticate their identities to nodes without disclosing private information. However, the protocol must also address the following concerns:

$\text{\textbf{(1).}}$ Users upload behavioral data from dynamic scenarios, and frequent authentication with nodes incurs additional time and computational overhead when interacting with different clusters.

$\text{\textbf{(2).}}$ Both users and nodes have limited computational resources, and lengthy interactions and parameter calculations during the identity authentication process can impact efficiency.

$\text{\textbf{(3).}}$ The authentication process should not be dominated by any single node to avoid the risk of a single point of failure.

This paper proposes cross-domain and continuous authentication to address these challenges. Users perform a single authentication with nearby nodes to establish legitimacy within the IoB. Additionally, a non-interactive ZKP protocol precomputes parameters, reducing the number of interactions required during the authentication process.

\subsection{Overview}

A portion of the nodes in the blockchain are set up as Certificate Authority (CA) nodes to manage user identities and nodes' credentials, and each user and CA node stores the {\textit{global variable}} needed for the ZKP protocol. Let $E({{F}_{p}})$ be an elliptic curve defined over a finite field ${{F}_{p}}$, where $p$ is a large prime. The {\textit{global variable}} is the base point $G$ on the curve that serves as a generator for the subgroup over $E({{F}_{p}})$ of prime order $n$. Due to the need to achieve the goal of user-behavior separation, users will treat their real identity as a secret, represented by the {\textit{private key}} $pr$, which is chosen uniformly at random from $[1,n-1]$. They compute a virtual identity using $G$ and $pr$ through elliptic curve scalar multiplication $pu=G\times [pr]$, and $pu$ is represented by a {\textit{public key}}, This {\textit{public key}} is then stored at a nearby CA node. During authentication, the CA node issues a {\textit{challenge}} to verify legitimacy, to which the user replies with a {\textit{response}} generated from their {\textit{private key}}, repeating the {\textit{challenge}}-{\textit{response}} process as necessary. In non-interactive protocols, the {\textit{challenge}} can also be generated by the user, reducing the number of interactions between the user and the node.

\subsection{Primary authentication}
When a user authenticates, the {\textit{challenge}} $c$ and {\textit{response}} $r$ will be computed simultaneously and their slices will be sent to multiple CA nodes via secret sharing.

\begin{list}{}{}

\item $\bullet$ {$\text{\textbf{User:}}$ The user randomly generates a variable $V$, which is calculated from a random number $v$ using $V=G\times [v]$. The {\textit{challenge}} is obtained by computing the hash digest $c=H(G||V||pu)$. The {\textit{response}} is calculated using $r=v-pr*c$.}

\item $\bullet$ {$\text{\textbf{Secret sharing:}}$ The user generates slices  $c\to \{{{c}_{1}},{{c}_{2}},...,{{c}_{q}}\}$ and  $r\to \{{{r}_{1}},{{r}_{2}},...,{{r}_{q}}\}$ for all CA nodes based on the original $c$ and $r$.}

\item $\bullet$ {$\text{\textbf{CA nodes:}}$ Upon receiving the $\{V,{{c}_{i}},{{r}_{i}}\}$ from the user, a CA node $i$ broadcasts its slice to reconstruct the {\textit{challenge}} and  {\textit{response}}. Finally, it performs verification by computing $\kappa =[r]\times G+pu\times [c]$. If $\kappa =V$, the validation passes.}
\end{list} 

Finally, we briefly analyze the computational cost. To prove the knowledge of the discrete logarithm for {\textit{public key}} $pu=G\times [pr]$ with base point $G$ over the elliptic curve, the user generates {\textit{proof}},{\textit{challenge}} and {\textit{response}}. The main cost is one scalar multiplication: $V=G\times [v]$. To verify the corresponding private key, the cost is approximately one multiplication over the elliptic curve: $\kappa =[r]\times G+pu\times [c]$.

\subsection{Continuous and multi-scenario authentication}

CA nodes reach a consensus on verification results to confirm consistency. After successful initial authentication, they attach a timestamp to the user's {\textit{public key}}, generating and storing a hash digest $H(publickey||timestamp)$ to indicate the validity period of the identity. The verifying CA node then issues $H(publickey||timestamp)$ for the user to keep. Importantly, users will not disclose it to protect personal information.

When the user moves from scenario A to scenario B, they will have to request permission to upload new behavioral data or access previously stored data. The user initiates an authentication request to the CA in scenario B, sending $H(publickey||timestamp)$ and their {\textit{public key}}. The CA checks whether the {\textit{public key}} has undergone initial verification and verifies its validity period by comparing the received $H(publickey||timestamp)$ with its stored hash value. If they match, the user is authenticated. If not, the CA node will reinitiate the verification process.

\section{Behavior processing layer}

The behavioral processing layer uses a consensus mechanism to process these data, which includes behavioral data, user identity, and node information. Given the limited computational resources of nodes, we employ PBFT to execute data consensus. This protocol determines the consensus result through information exchange among nodes, ultimately based on the majority of votes. By utilizing a voting system for consensus, the PBFT algorithm does not require computation-intensive problems (e.g., POW), providing a lightweight consensus solution for blockchain nodes. However, this approach has some drawbacks in IoB, particularly in scalability. Following the previous example, the user will not always be in the company-restaurant-home scenario, when the user has a new activity area and generates new behavior, nearby nodes will join the blockchain to process the behavioral data. This is not an accidental situation, but one that is growing in real time and at a rapid pace. However, the original PBFT protocol requires extensive message exchanges among nodes, making communication and computational overhead untenable when the number of nodes in the chain grows quickly \cite{WW14}. This paper proposes clustering nodes based on location and communication capabilities. Nodes within a cluster should have similar communication capabilities to avoid excessive communication delays, while geographic proximity ensures lower data upload latency for users.

\subsection{Node clustering}

Density-based spatial clustering of applications with noise (DBSCAN) is a density-based clustering algorithm that groups objects in dense regions into clusters, capable of discovering arbitrary cluster shapes. Traditional DBSCAN requires two inputs: a user-defined radius (EPS) and the minimum number of edge points (MinPts) within that radius. This paper considers both the geographical location and communication capabilities of nodes, necessitating two-dimensional data. Therefore, we employ a multi-dimensional attribute DBSCAN approach, incorporating two distance metrics: EPS1, which measures the Euclidean distance for geographical proximity, and EPS2, which assesses the similarity of communication capabilities. EPS1 and EPS2 represent spatial and non-spatial attributes, respectively, and can be measured using either Euclidean or Manhattan distances. MinPts denotes the minimum number of nodes within the distance of EPS1 and EPS2 for a given node. As with the DBSCAN algorithm, we traverse each node to find out all the core objects, which need to satisfy that the number of neighboring nodes at the intersection of EPS1 and EPS2 is greater than MinPts, and these connected core objects will form clusters.

\subsection{Data consistency test}

The nodes in the group are categorized as leader, secondary, and CA nodes. Leader and secondary nodes execute the consensus protocol and store user-uploaded data. Users randomly select a nearby node as the leader node to ensure real-time information transmission. All nodes save each other's public keys, IP addresses, and status information. If a node's management information differs from that in the CA node, it can request a consistency check, prompting all nodes to synchronize their data.
When users upload new behavioral data, they send a request with timestamps to the leader node, which uses the PBFT three-phase protocol to achieve consensus, ensuring data consistency across the cluster. Each node includes its signature and a hash digest when forwarding messages to maintain message integrity and sender legitimacy.

\subsection{Intergroup data consistency test}

The DBSCAN algorithm clusters nodes with similar communication capabilities and geographical locations. This leads to significant variability in communication capabilities among nodes in inter-cluster consensus, making the PBFT algorithm less effective. To address this, we use a gossip-like protocol for inter-cluster consensus, where the leader node selects peer nodes to receive new block messages. These leader nodes then relay the messages until all nodes in the network are informed. A key challenge is the time it takes for the new block to reach all leader nodes. The selection of relay nodes plays a crucial role in enhancing message propagation efficiency. Like the intra-cluster PBFT approach, we consider the geographical location and communication capabilities of relay nodes to increase the likelihood of leader nodes choosing them.

First, the leader node $Mnode$ that generates a new block confirms the completion of intra-cluster consensus. For the remaining $k$ leader nodes within a distance of EPS3, the Euclidean distance ${{l}_{1}},{{l}_{2}}...{{l}_{k}}$ is calculated to obtain the average distance $\overset{-}{\mathop{l}}\,$. The $Mnode$ then sends acknowledgment messages to these $k$ leader nodes, recording the time difference ${{\lambda }_{1}},{{\lambda }_{2}},...{{\lambda }_{k}}$ between the replies and the sent acknowledgments, and get the average time difference $\overset{-}{\mathop{\lambda }}\,$. Weights are assigned based on the ratio of each node's distance and time difference to the average. Nodes with shorter distances and smaller time differences have higher weights, making them more likely to be chosen for message forwarding.

\section{Behavior management layer}

The behavioral management layer provides data storage and access functions to ensure that legitimate users have secure access to behavioral data to prevent data leakage and tampering. This layer must implement access control and data sharing for users with different permissions, allowing them to access their stored behavioral data and other data within the same behavioral category for sharing purposes. However, access to data from unrelated behavioral categories should be restricted to protect user privacy. Therefore, an access control system can be established to support personalized data access and sharing for users with the same permissions. Additionally, this system should integrate with distributed systems to facilitate collaborative permission verification among nodes.

\begin{table}[htbp]
\renewcommand{\arraystretch}{2}
\centering

\caption{Access Control List}

\resizebox{\columnwidth}{!}{%
\begin{tabular}{|c|c|c|c|c|} \hline
Behavior Category & Sports & Driving & ... & Going Home  \\ \hline
DPF & $i=1, p_{i}(i)=1$   & $i=2, p_{i}(i)=1$     &... &$i=N, p_{i}(i)=1$               \\ \hline
Verification keys  &$g^{\partial_{1}}$  &$g^{\partial_{2}}$  &...  & $g^{\partial_{N}}$                                     \\ \hline
Access keys  &$\partial_{1}$  &$\partial_{2}$  &... &$\partial_{N}$                          \\ \hline
\end{tabular}%
}
\end{table}

\subsection{Access control list}

By establishing a one-to-one correspondence between different types of data and their associated permissions, access control lists can enable users to access data with varying levels of permissions. From a cryptographic perspective, the data access control list models different types of behavior corresponding to keys. Each behavior database is associated with a verification key ($vk$) and an access key ($sk$). Therefore, this list associates behavior with verification keys, providing access rights to users who possess the access key. Notably, the specific database that users request access to is hidden from distributed nodes, preventing privacy breaches.

\subsection{Function secret sharing}

FSS can provide privacy protection for distributed systems by allowing multiple nodes to collectively verify user permissions without exposing user access requirements. In FSS, a user shares a function ${{f}_{i}}$ with distributed node $j,1\le j\le s$, and the secret share of the function is represented as the function key ${{k}_{j}}$ or share $\left[ {{f}_{i}} \right]$, which does not leak any information about ${{f}_{i}}$. A set of function secret shares ${{k}_{j}}$ for an input x can recover ${{f}_{i}}(x)$. Distributed Point Functions (DPF) are one of the main primitives for constructing FSS. Inspired by \cite{WW15}, we establish a connection between DPF and multiple behavioral databases. The DPF ${{P}_{i}}$ outputs a specific value $m$ when the input is $i$, and outputs 0 when the input is not equal to $i$. Specifically, the DPF secret share for a node with function key ${{k}_{j}}$ when inputting $i$ is ${{[{{P}_{i}}]}_{j}}\leftarrow DPF.Eval({{k}_{j}},i)$, and summing the secret shares $[{{P}_{i}}(i)]$ for the input $i$ can recover $m={{P}_{i}}(i)$.

In a database that stores $N$ behavioral categories, the access key for user access to the corresponding behavioral data is defined as $sk={{\partial }_{k}},1\le k\le N$. For simplicity, let $m=1$, allowing us to construct the verification key $v{{k}_{1}},...,v{{k}_{N}}$ for the access control list $\Upsilon =(v{{k}_{1}},...,v{{k}_{N}})$ as ${{g}^{{{\partial }_{1}}}},...,{{g}^{{{\partial }_{N}}}}$, where $v{{k}_{N}}$ corresponds to the point function ${{P}_{N}}=({{0}_{1}},{{0}_{2}}...,{{1}_{N}})$. For a specific behavioral database associated with ${{\partial }_{k}}$ and ${{g}^{{{\partial }_{k}}}}$, when node $j$ receives the function key ${{k}_{j}}$, it retrieves using input values $1,...,N$ and ${{g}^{{{\partial }_{1}}}},...,{{g}^{{{\partial }_{N}}}}$, to obtain the corresponding secret share of the verification key ${{g}^{{{\left[ {{\partial }_{k}} \right]}_{j}}}}$ from the database. To prove ownership of ${{\partial }_{k}}$, which corresponds to the access rights, the user generates a proof $\pi :=-{{\partial }_{k}}$ and distributes the secret shares $\left[ \pi  \right]$ to the distributed nodes. Each node can locally execute the verification ${{g}^{{{[{{\partial }_{k}}]}_{j}}}} \times  {{g}^{{{[\pi ]}_{j}}}} $ to obtain ${{\tau }_{j}}:={{g}^{{{[0]}_{j}}}}$. All nodes can ultimately verify $\tau ={{g}^{0}}$ to confirm access rights to the behavioral data $k$.

\subsection{Private data access control}

Personalized privacy access control establishes an access control list $\Upsilon =(v{{k}_{1}},...,v{{k}_{N}})$ by mapping each category of behavioral data to different input values of DPF, as shown in Table \uppercase\expandafter{\romannumeral1}. When a user with an access key requests access to a specific category of data, secret shares are generated and distributed to $s$ nodes. The distributed nodes collectively verify whether the user has access rights to the function without disclosing the associated behavioral category of the function. The verification process is shown in Algorithm 1 and consists of three components: KeyGen, LocalVerify, and CheckAccess.

\begin{list}{}{}
\item $\bullet$ {$\text{\textbf{KeyGen: }}$ Using $i=k,p{}_{i}(i)=1$ and the security parameter, a trusted third-party organization outputs a pair of access keys and verification keys $(sk,vk)$. The verification keys $v{{k}_{i}}$, associated with different behavioral category databases, form the access control list $\Upsilon =(v{{k}_{1}},...,v{{k}_{N}})$, which is stored by the participating nodes. The access keys are held by the user, who distributes $s$ function keys ${{k}_{1}},{{k}_{2}},...,{{k}_{s}}$ and generates proof $\pi :=-{{\partial }_{k}}$ using the access key, distributing $s$ proof secret shares ${{[\pi ]}_{1}},{{[\pi ]}_{2}},...,{{[\pi ]}_{s}}$ to the distributed nodes.}

\item $\bullet$ {$\text{\textbf{Localverify: }}$ Distributed node $j$ obtains the verification key multiplicative secret share ${{g}^{{{\left[ {{\partial }_{k}} \right]}_{j}}}}$ related to the user's access key based on the access control list $\Upsilon $ and function key ${{k}_{j}}$. Combining this with the received proof secret share ${{[\pi ]}_{j}}$, node $j$ locally computes ${{\tau }_{j}}:={{g}^{{{[0]}_{j}}}}$.}

\item $\bullet$ {$\text{\textbf{CheckAccess: }}$ The distributed nodes collectively input the local verification share ${{\tau }_{j}}$ to validate $\tau ={{g}^{0}}$, confirming whether the user has access rights to the function.}
\end{list}

\renewcommand{\algorithmicrequire}{\textbf{Input:}}  
\renewcommand{\algorithmicensure}{\textbf{Output:}} 

\begin{algorithm}[h]
  \caption{ Private data access control} 
  \label{alg::conjugateGradient}
  \begin{algorithmic}[1]
    \Require
      Security parameter, distributed point functions, and integers $2\le t\le s$;
	\Ensure
		Verification result $\tau$;

 \Statex\textbf{KeyGen:} 
    \State $v{{k}_{k}}={{g}^{{{\partial }_{k}}}}$, $s{{k}_{k}}={{\partial }_{k}}$
 \State $({{[\pi ]}_{1}},{{[\pi ]}_{2}},...,{{[\pi ]}_{s}})\leftarrow Share(sk)$
\State  return $(v{{k}_{k}},s{{k}_{k}})$, $({{[\pi ]}_{1}}, {{[\pi ]}_{2}},...,{{[\pi ]}_{s}})$

 \Statex\textbf{Localverfy$(\Upsilon ,{{k}_{j}},{{[\pi ]}_{j}})$:}
\State parse $\Upsilon =(v{{k}_{1}},...,v{{k}_{N}})$ and ${{k}_{j}}$
\State ${{[{{p}_{i}}]}_{j}}\leftarrow DPF.Eval({{k}_{j}},i)$, $\forall i\in \{1,...,N\}$
\State btain a secret share ${{g}^{{{[{{\alpha }_{k}}]}_{j}}}}:=\prod\nolimits_{i=1}^{N}{{{({{g}^{{{\alpha }_{i}}}})}^{{{[{{p}_{i}}]}_{j}}}}}$
\State ${{\tau }_{j}}:={{g}^{{{[0]}_{j}}}}\leftarrow {{g}^{{{[{{\partial }_{k}}]}_{j}}}}{{g}^{{{[\pi ]}_{j}}}}$
\State return ${{\tau }_{j}}$

 \Statex\textbf{CheckAccess:} 
\State parse {${{\tau }_{1}},...{{\tau }_{t}}$}
\State	All verifiers proceed to check that $\tau ={{g}^{0}}$
\State return $\tau \overset{?}{\mathop{=}}\,1$
  \end{algorithmic}
\end{algorithm}

\section{Evaluation and Result Analysis}     
 To simulate nodes in various geographic locations, this paper uses the Gowalla dataset, which contains 6,442,892 location-based check-in records. After processing, latitude and longitude information is retained, and an additional dimension for communication capability is added to each record. We set the user activity area in Beijing, with latitude 39.433333 to 41.05 and longitude 115.416666 to 117.50, to get 5,349 nodes. Using DBSCAN clustering shown in Fig. 3, these nodes are classified into 23 clusters. This clustering significantly reduces the number of nodes needed for PBFT consensus.

\begin{figure}[!t]
\centering
\includegraphics[width=4.0in]{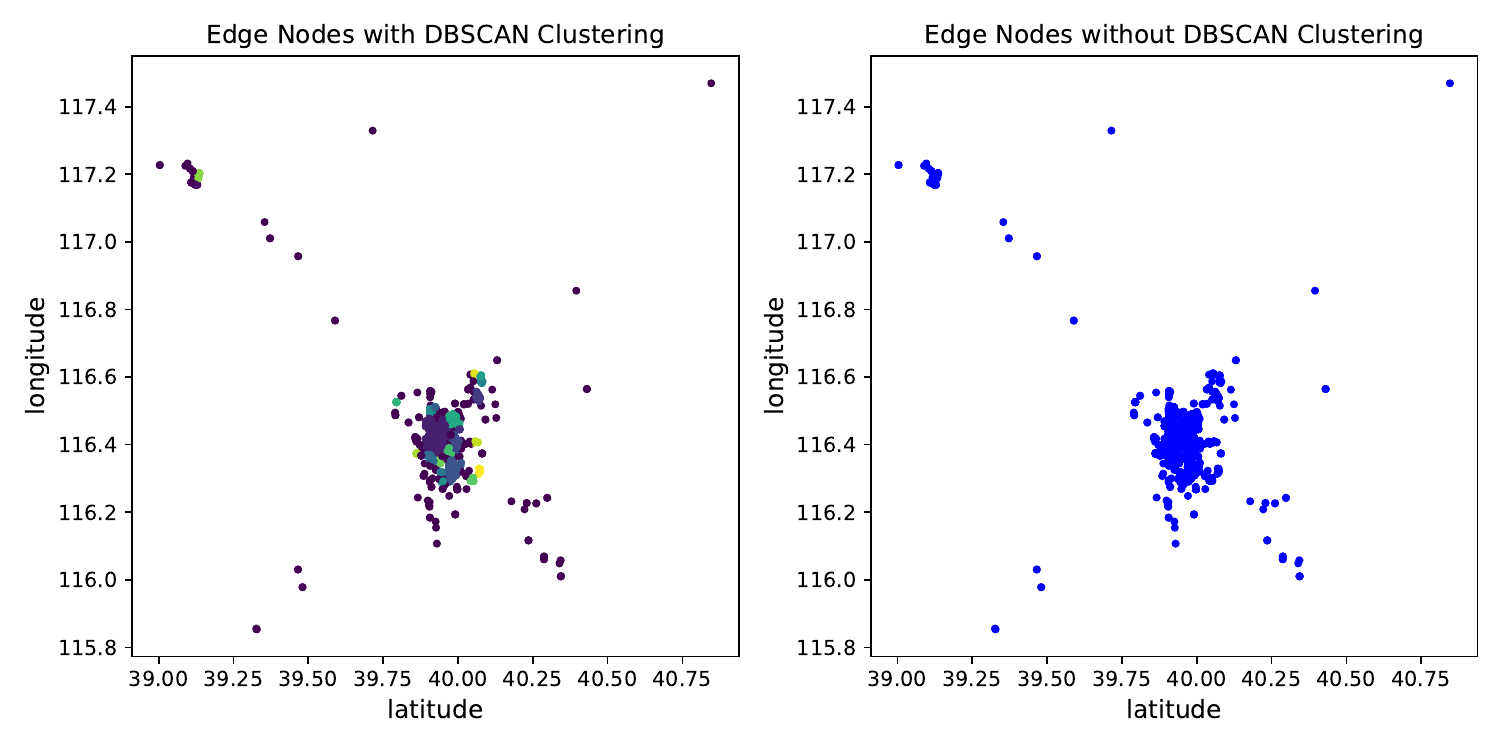}
\caption{DBSCAN Clustering}
\label{Fig3}
\end{figure}

\begin{figure}[!t]
\centering
\includegraphics[width=3.0in]{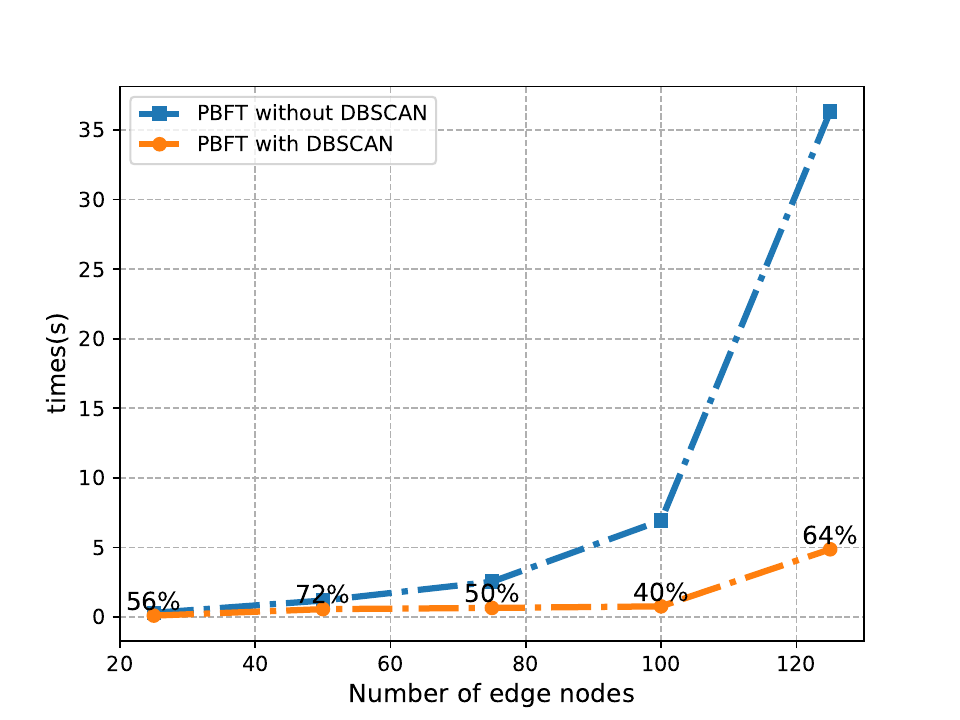}
\caption{Performance of consensus time}
\label{Fig4}
\end{figure}

The number of nodes executing PBFT consensus directly affects network performance, such as consensus time. By using DBSCAN clustering, nodes with varying geographic locations and communication capabilities can be clustered, improving consensus efficiency. The communication complexity of PBFT is generally approximated as $O({{N}^{2}})$. When the number of nodes is reduced, e.g. by 60\%, actual communication cost is only 0.16 times the original cost. To further demonstrate the effectiveness of the approach, we first select 25 to 125 nodes from the area and perform clustering using DBSCAN, followed by running the PBFT algorithm. As shown in Fig.4, after clustering, the maximum number of nodes executing PBFT is only 40\% to 72\% of the original node count, significantly reducing the consensus time, with an average decrease of 74\%.

\begin{figure}
	\centering
	\subfloat[]{
	\includegraphics[width=3.0in]{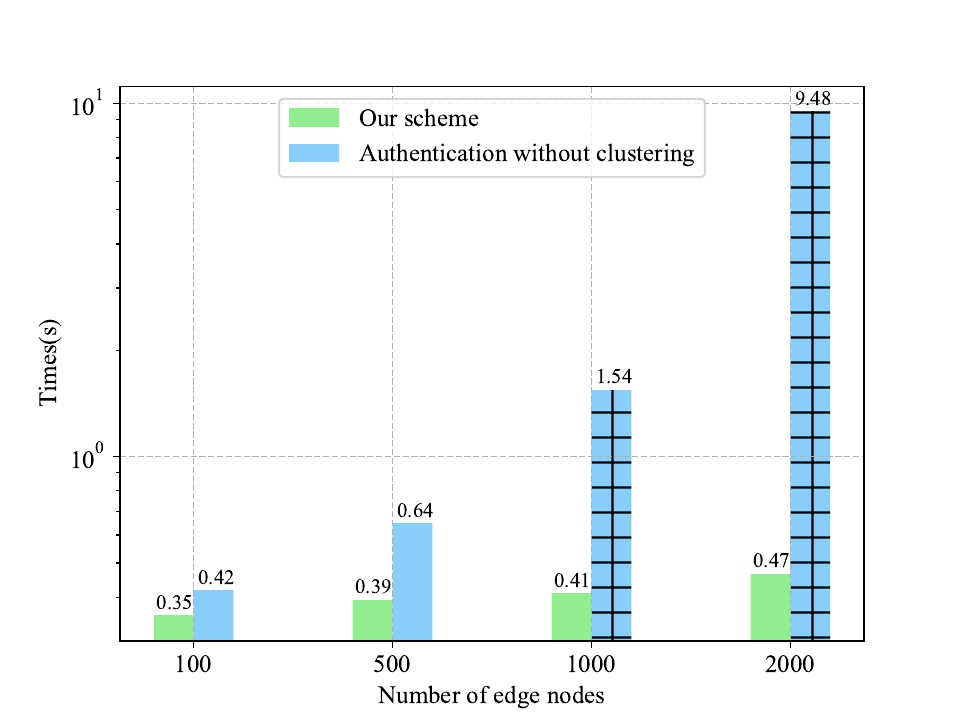}} 
   \quad  
    \subfloat[]{
	\includegraphics[width=3.0in]{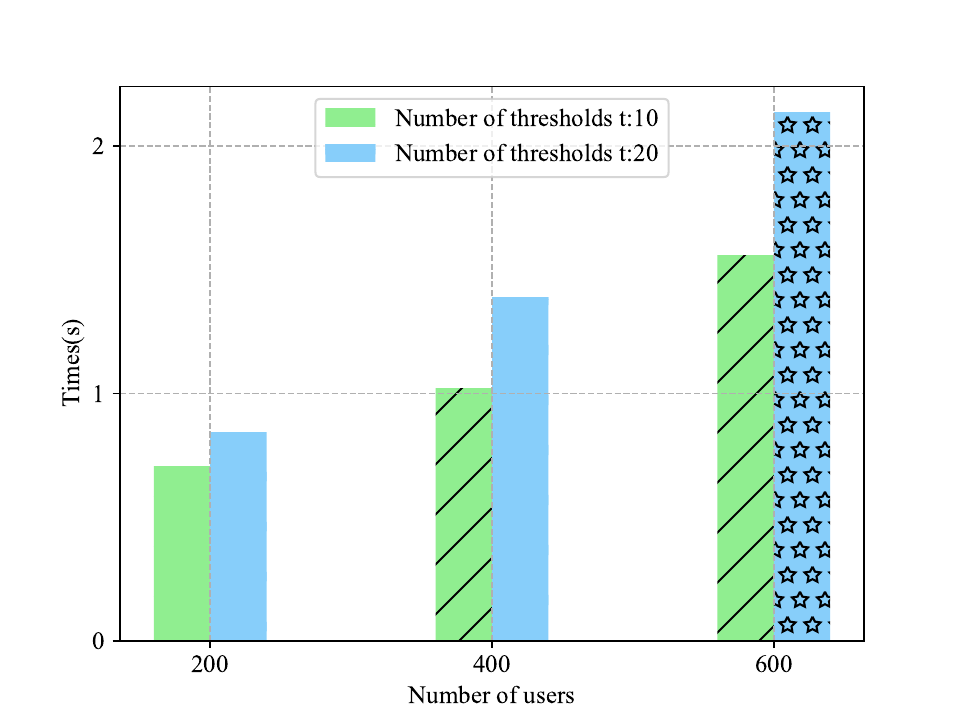}}
	\caption{Performance of authentication cost: a) authentication w/ and w/o clustering; b) authentication w/ different number of users.}
	\label{fig5}
\end{figure}

Fig.5 illustrates the time cost for users to perform authentication. To validate the effectiveness of the proposed approach in the behavior processing layer, Figure 5(a) tests different node counts, setting the secret recovery threshold to 1/4 of the total. Clustering of nodes effectively reduces the number of CA nodes, keeping authentication time nearly constant as the count increases from 100 to 2,000, with an average time reduction of 56\%. Without clustering, secret shares must be distributed to each node, increasing the number of nodes executing secret recovery. When nodes reach 2,000, authentication time rises from 0.47 to 9.48 seconds, indicating good scalability for multi-scenario IoB services in our scheme. Then, this paper verifies the feasibility of multiple users simultaneously performing authentication. As user requests increase, the frequency of secret recovery by nodes rises, but overall time remains manageable. Figure 5(b) confirms that the proposed solution effectively meets the authentication needs of multiple users.

\begin{figure}[!t]
\centering
\includegraphics[width=3.0in]{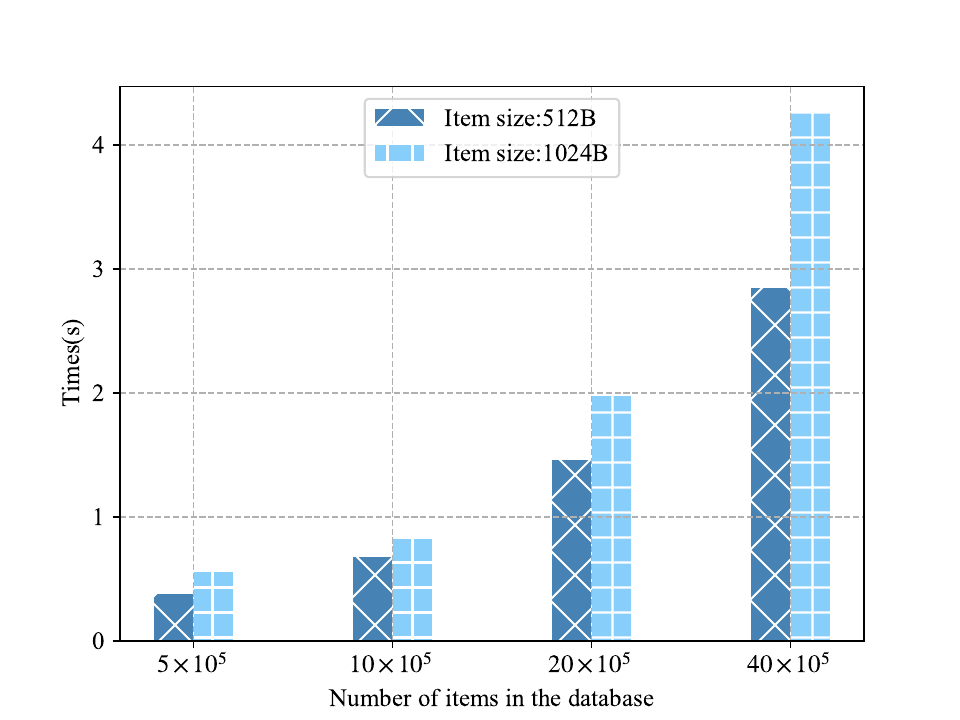}
\caption{Performance overhead of the access control}
\label{Fig6}
\end{figure}
In Fig. 6, we fixed the item size to 512B and 1024B and tested the overhead of privacy access control. We show that the overhead of privacy access control increases with the number of behavior database items, but generally stays within an acceptable range, which indicates the feasibility of the privacy protection mechanism of the proposed scheme.

\section*{CONCLUSION}

This paper proposes a Blockchain-based distributed IoB. We consider the security and privacy risks under each stage (sensing, uploading, and accessing) of the data with dynamically changing characteristics of user scenarios.We decompose the framework into a multi-layer architecture to achieve data management under multi-scenario IoB through intra-layer stage division and inter-layer joint analysis. On the one hand, we fully consider the features of dynamic switching of user activity regions, and successfully realize the effective combination of authentication and consistency checking with blockchain by improving the traditional ZKP into an authentication scheme that supports cross-domain and further integrating node clustering with geographic and computational differences. On the other hand, we introduce the viewpoint of same-privilege sharing in access control, and design a secure and effective data sharing mechanism under distributed architecture through FSS. Simulation results show that the proposed framework can effectively reduce the consensus and authentication time.

However, here are some challenges that need to be addressed in future work.First, we effectively improve the consensus effectiveness after the growth of the number of nodes by clustering. However, unlimited growth of nodes and concurrent uploading of data from multiple users are the current challenges. Therefore the next plan is to consider cloud-edge hierarchical architecture and study the effective combination of cloud computing and decentralized data architecture.Second, we have not yet effectively explored security during data transmission. In future work, we will explore how to apply physical layer security and AI to address data transmission policies under multi-scenario IoB.

\bibliographystyle{IEEEtran} 
\bibliography{wenxian}	

\section*{BIOGRAPHIES}
\vspace{-1cm}
\begin{IEEEbiographynophoto}{Shiqi Zhang}
is currently pursuing the Ph.D. degree in information and communication engineering with the School of Communications and Information Engineering, Chongqing University of Posts and Telecommunications, Chongqing, China. His research interests include the network security and privacy, data markets and edge computing.
\end{IEEEbiographynophoto}
\vspace{-1cm}
\begin{IEEEbiographynophoto}{Dapeng Wu (Senior Member, IEEE)}
received his Ph.D. degree in 2009 from the Beijing University of Posts and Telecommunications, Beijing, China. Now, he is a professor at the Chongqing University of Posts and Telecommunications, Chongqing, China. He is the inventor and co-inventor of 28 patents and patent applications. His research interests are in social computing, wireless networks, and big data. Prof. Wu serves as TPC Chair of the 10th Mobimedia and program committee member for numerous international conferences and workshops. He served or is serving as an Editor or/and Guest Editor for several technical journals, such as IEEE Internet of Things Journal, Elsevier Digital Communications and Networks, and ACM/Springer Mobile Network and Applications. He is an IEEE senior member.
\end{IEEEbiographynophoto}
\vspace{-1cm}
\begin{IEEEbiographynophoto}{Honggang Wang (Fellow, IEEE)}
 is currently the founding Chair and Professor with the Department of Graduate Computer Science and Engineering, Katz School of Science and Health, Yeshiva University, New York, NY, USA. He was a Professor with UMass Dartmouth, Dartmouth, MA, USA. He is an alumnus of NAE Frontiers of Engineering program. He graduated 30 MS/Ph.D. students and produced high-quality publications in prestigious journals and conferences in his research areas, winning several prestigious best paper awards. His research interests include Internet of Things and its applications in health and transportation domains such as
autonomous vehicles, machine learning and Big Data, multimedia and cyber security, smart and connected health, wireless networks, and multimedia communications. He is an IEEE distinguished Lecturer and a Fellow of AAIA. He was the Editor in Chief of IEEE INTERNET OF THINGS JOURNAL during 2020–2022. He was the past Chair during 2018–2020 of IEEE Multimedia Communications Technical Committee and past IEEE eHealth Technical Committee Chair during 2020–2021.
\end{IEEEbiographynophoto}
\vspace{-1cm}
\begin{IEEEbiographynophoto}{Ruyan Wang}
received his Ph.D. degree in 2007 from the University of Electronic and Science Technology of China, Sichuan, China. He is the Dean of the School of Communications and Information Engineering, Chongqing University of Posts and Telecommunications. He is the recipient of the Danian Huang Team from the Ministry of Education of the Peoples Republic of China. His research interests include network performance analysis and multimedia information processing.
\end{IEEEbiographynophoto}

\vfill

\end{document}